\documentclass{emulateapj}

\usepackage{times}   
\usepackage{graphicx}
\usepackage{natbib}
\usepackage{sidecap}   

\newcommand{\kms}{km~s$^{-1}$\,}
\newcommand {\sm}{\rm\,M$_\odot$}

\shorttitle{Kinematic status and mass content of the Sculptor dSph}
\shortauthors{Battaglia et al.}

\begin{document}

\defcitealias{tolstoy2004}{T04}


\title{The kinematic status and mass content of the Sculptor dwarf spheroidal galaxy\altaffilmark{1}}

\altaffiltext{1}{Based on FLAMES observations collected at the ESO, 
proposals 171.B-0588 and 076.B-0391A}


%
%
%

\author{G.~Battaglia\altaffilmark{2,3}, A.~Helmi\altaffilmark{3}, E.~Tolstoy\altaffilmark{3}, M.~Irwin\altaffilmark{4}, 
V.~Hill\altaffilmark{5}, P.~Jablonka\altaffilmark{6}}
\altaffiltext{2}{Current address: European Organization for Astronomical Research in the Southern Hemisphere; K. Schwarzschild-Str. 2, 85748 Garching, Germany; gbattagl@eso.org}
\altaffiltext{3}{Kapteyn Astronomical Institute, University of Groningen, P.O.Box 800, 9700 AV Groningen, the Netherlands; ahelmi, etolstoy@astro.rug.nl}
\altaffiltext{4}{Institute of Astronomy, Madingley Road, Cambridge CB03 0HA, UK; mike@ast.cam.ac.uk}
\altaffiltext{5}{GEPI, Observatoire de Paris, CNRS, Universit\'e Paris Diderot ; Place Jules Janssen 92190 Meudon, France; Vanessa.Hill@obspm.fr}
\altaffiltext{6}{Observatoire de Gen\`eve, Laboratoire d'Astrophysique de
l'Ecole Polytechnique F\'ed\'erale de Lausanne (EPFL), CH-1290 Sauverny, Switzerland; Pascale.Jablonka@obs.unige.ch}

%


\begin{abstract}
We present VLT/FLAMES spectroscopic observations (R$\sim$6500) 
in the \ion{Ca}{2} triplet 
region for 470 probable kinematic members of the Sculptor (Scl) dwarf spheroidal galaxy. 
The accurate velocities ($\pm$2 \kms) 
and large area coverage of Scl allow us to measure a velocity gradient of 7.6$_{-2.2}^{+3.0}$ \kms deg$^{-1}$ 
along the projected major axis of Scl, likely  a signature of intrinsic rotation. 
We also use our kinematic data to
measure the mass distribution within this system. By considering independently the kinematics of 
the two distinct stellar components known to be present in Scl, we are able to relieve known degeneracies, and find that 
the observed velocity dispersion 
profiles are best fitted by a cored dark matter halo with core radius $r_c=$ 0.5 kpc and 
mass enclosed within the last measured point $M(< 1.8 \mathop{\rm kpc})=3.4 \pm 0.7 \times 10^8$ \sm, assuming an increasingly 
radially anisotropic velocity ellipsoid. This results in a mass-to-light ratio of 158$\pm$33 (M/L)$_{\odot}$ 
inside 1.8 kpc. An NFW profile with concentration $c=20$ and mass $M(< 1.8 \mathop{\rm kpc}) = 2.2_{-0.7}^{+1.0} \times 10^8$ \sm 
is also statistically consistent with the observations, but it tends to yield poorer fits for the metal rich stars. 
\end{abstract}





\section{INTRODUCTION}
Recent wide area photometric and spectroscopic surveys of large
numbers of stars in Local Group dwarf spheroidal galaxies (dSphs) have
resulted in major progress in our knowledge of these objects and a
more complete picture of their characteristics.  They appear to be the
most dark matter (DM) dominated objects known to date, with
mass-to-light (M/L) ratios larger than 100 \citep[e.g.,][and
references therein]{kleyna2005, walker2007}; their kinematic status is
generally believed to be undisturbed by interactions with the Milky
Way (MW), with the exception of Carina \citep[see][and references
therein]{munoz2006car}, Bo\"otes \citep[e.g.,][]{belokurov2006}, and
perhaps Ursa Minor \citep[e.g.,][]{bellazzini2002}.  In some cases a
number of spatial and kinematic substructures have been observed
\citetext{e.g., \citealt{wilkinson2004}; \citealt{coleman2005};
\citealt{battaglia2007}, hereafter B07}, which are perhaps the
remnants of accreted subunits (stellar clusters or smaller galaxies).
Furthermore, the combination of metallicity ([Fe/H]) and kinematic
information have unveiled the presence of multiple stellar components
in the Sculptor (Scl) and Fornax dSphs \citetext{\citealt[][hereafter
T04]{tolstoy2004}; \citealt{battaglia2006}}, where the metal rich (MR)
stars have a more centrally concentrated, less extended spatial
distribution, and show colder kinematics than the metal poor (MP)
stars.

The presence of multiple stellar populations may require a
modification in the way these systems are dynamically modeled to
derive their mass content.  Traditionally, their mass distribution has
been derived from the Jeans analysis of the line-of-sight (l.o.s.)
velocity dispersion obtained considering all stars as a single
component embedded in an extended DM halo. This analysis is subject to
the well known degeneracy between the mass distribution and the
orbital motions of the individual stars in the system (mass-anisotropy
degeneracy), which prevents a distinction amongst different DM
profiles.

In this Letter we show that it is possible to partially break this
degeneracy by modeling Scl as a two-(stellar)component system
embedded in a DM halo.  As part of the modeling of Scl, we also
analyzed its dynamical status, and for the first time, we determine
the presence of (statistically significant) intrinsic rotation in the
stellar component of a dSph.

\section{OBSERVATIONS}

We acquired photometric data from the ESO/2.2m Wide Field Imager (WFI)
at La Silla, and spectroscopic follow-up data in the Ca~II triplet
(CaT) region from FLAMES at the VLT. Our spectroscopic targets were
chosen from the WFI imaging to have the colors and magnitudes
consistent with Red Giant Branch (RGB) stars. We acquired data for 15
different fields in Scl (7 fields in addition to those presented in
T04). For details of the data reduction and analysis see
\citet{battaglia2006, battaglia2008}. From the imaging we obtained the
surface density profile of the different stellar populations in Scl
and from the spectroscopy we measured line-of-sight (l.o.s.)
velocities, and CaT metallicities ([Fe/H]) calibrated using the 
relation derived in \citet{battaglia2008}.

Using an iterative procedure we measure a systemic velocity $v_{\rm
hel, sys} = 110.6 \pm 0.5$ \kms and a dispersion $\sigma = 10.1 \pm
0.3$ \kms for Scl. A simple kinematic membership selection (stars with
heliocentric l.o.s.\ velocity within 3$\sigma$ from $v_{\rm hel,
sys}$) produces 470 probable Scl members, as shown in
Fig.~\ref{fig:rotation}a.  We use this sample when investigating the
kinematic status of Scl (Sect.~\ref{sec:kinematicstatus}), and adopt a
more accurate membership criterion when deriving the l.o.s.\
dispersion profile (Sect.~\ref{sec:dispersion}).

\section{KINEMATIC STATUS AND ROTATION} \label{sec:kinematicstatus}

We first focus on the kinematic status of Scl in terms of tidal
disruption and intrinsic rotation, as both these factors can inflate
the observed l.o.s.\ velocity dispersion profile.

\begin{figure}
\begin{center}
\plotone{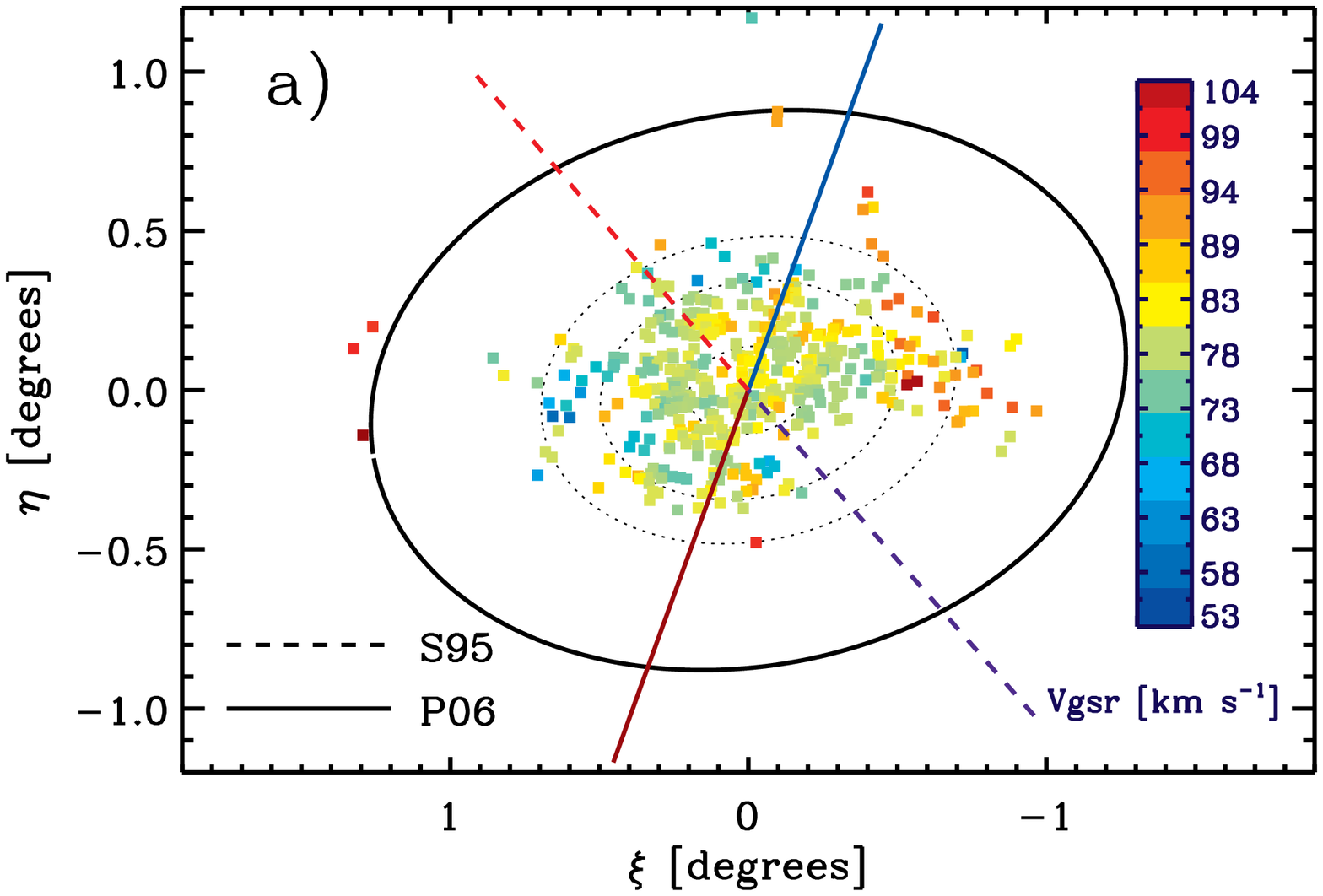}
\plotone{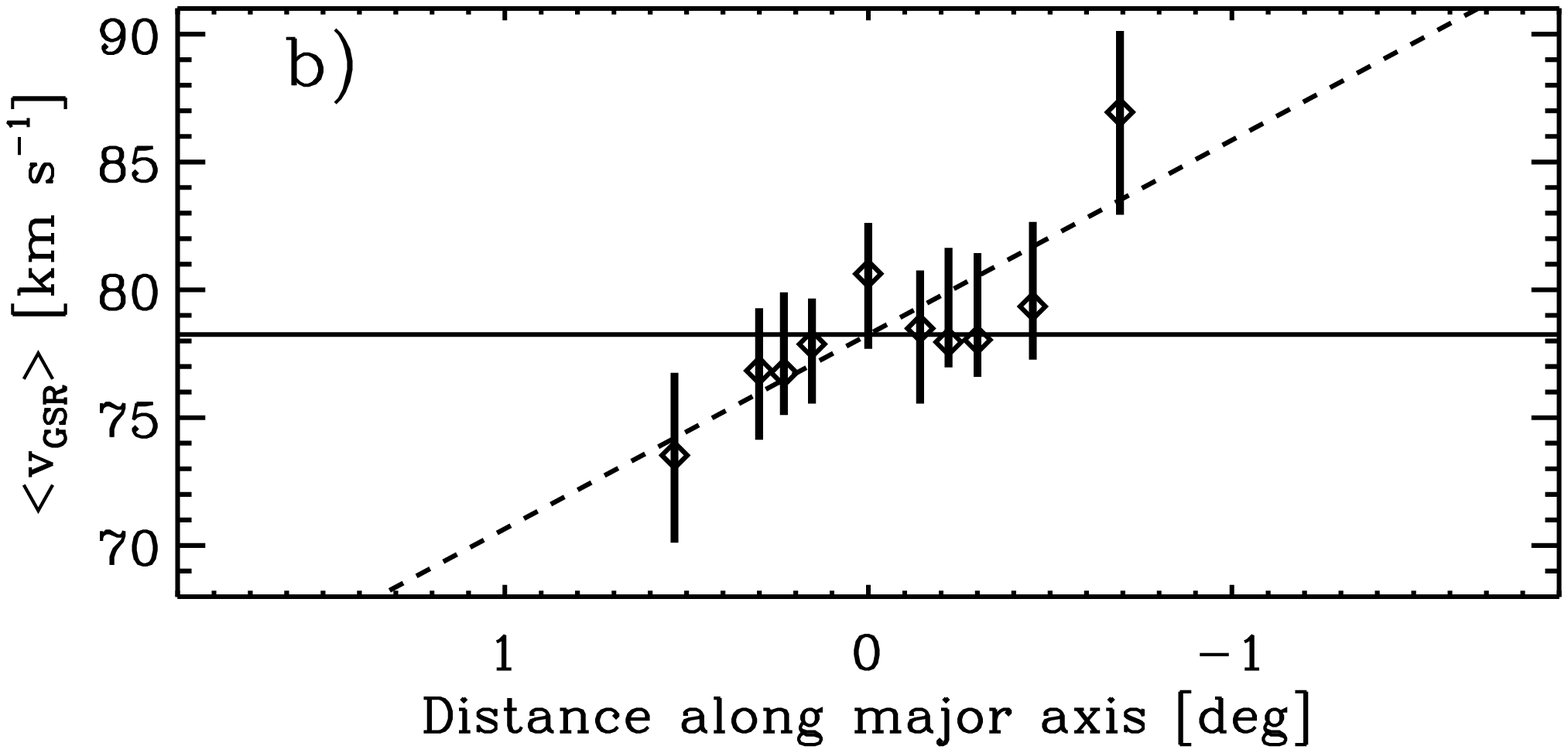}
\caption{a) Velocity field, smoothed with a median filter, for
3$\sigma$ Scl kinematic members. We use the velocities in the GSR
frame to avoid spurious gradients introduced by the component of the
Sun and Local Standard of Rest motion's along the l.o.s.\ to Scl.  The color bar gives
the velocity scale. The dotted ellipses are placed at 0.2, 0.5, 0.7
deg to give an idea of the distance scale. The solid ellipse shows the
nominal tidal radius \citep[from][]{ih1995}.  The dashed and solid
lines indicate, respectively, Scl's orbit calculated using the proper
motion measured by \citet{schw1995} and \citet{piatek2006}, where red
and blue shows the receding and approaching velocities.  b) Median
$v_{\rm GSR}$ per bin versus distance along the projected major axis
(diamonds with error-bars) for 3$\sigma$ Scl kinematic members located
within $\pm 0.15$ deg from this axis.  Each bin contains $\sim$30
stars. The solid and dashed lines show, respectively, Scl systemic
velocity and the best fit to the observed velocity gradient (see
text). Positive distances correspond to the east side of the galaxy,
negative distances to the west side.}
\label{fig:rotation}
\end{center}
\end{figure}

No signs of tidal disruption such as tidal tails and S-shaped contours
are found in our photometric data-set (Battaglia et al., in
prep), although they are not necessarily
expected as shown in N-body simulations by \citet{munoz2008}.  The
presence of a velocity gradient along the direction of the galaxy's
orbital motion (indicated by the proper motion direction) is another
possible sign of tidal disruption \citep[e.g.][]{oh1995}.  We will
call such a velocity gradient ``apparent rotation'' to distinguish it
from intrinsic rotation. Fig.~\ref{fig:rotation}b shows the Galactic
standard of rest (GSR) velocities of the probable members of Scl found
within $\pm$0.15 deg from its projected major axis (position angle
PA$=99^{\circ}$). At a projected radius $R\sim-0.7$~deg the median
velocity is $8.7_{-4.0}^{+3.1}$ \kms larger than the systemic and at
$R\sim+0.5$~deg it is $4.7_{-3.4}^{+3.2}$ \kms lower.  The
best-fitting straight-line gives a velocity gradient of
$7.6_{-2.2}^{+3.0}$ \kms deg$^{-1}$.  For comparison, no velocity
gradients are found along the minor axis (PA$=9^{\circ}$) and other
two explored axes at intermediate PA (45$^{\circ}$, 135$^{\circ}$).
Other techniques such as bisection \citep{walker2006} support these
results. This signal is unlikely to be significantly affected by
the foreground since at a projected distance from Scl's center of 0.5-0.7 deg 
the expected fraction of
MW contaminants in the 3$\sigma$ membership range is only $\sim$10\%.

If the measured velocity gradient is caused by tidal disruption, then
it should arise along the direction of Scl's proper motion, which
appears not to be the case according to the measurements by
\cite{schw1995} and \citet{piatek2006}. Although the errors on the proper motions
are large, it is encouraging that both works agree on the amplitude
along the North-South direction, indicating that it is unlikely that
more accurate measurements will lead to an orbit aligned with the
major axis of Scl.  In Fig.~1a we also show the resulting orbits
obtained by integrating these proper motions and the l.o.s.\ velocity
of Scl in a Galactic potential. If the tidal debris is aligned with
the orbit, then approaching velocities would be detected on the East
side of the galaxy, and receding on the West side, opposite to what is
observed. Furthermore, the simulations by \cite{munoz2008} also show
that velocity gradients due to tidal disruption usually appear at
larger projected radii, equal to the King limiting radius ($\sim$ 1.3
deg for Scl, see Irwin \& Hatzidimitriou 1995) with amplitudes of
$\lesssim$2 \kms, much smaller than measured here. From these
arguments we conclude that it is more likely that the detected
velocity gradient is due to intrinsic rotation than to tidal
disruption \citep[see][for a different interpretation based mostly on 
photometric data]{westfall2006}. This would therefore be the first time that statistically
significant rotation has been found in the stellar component of a
dSph.

In the following analysis we use rotation-subtracted GSR
velocities. The rotational velocity subtracted to each star is $v_{\rm
rot}= kx$, where $k = -7.6_{-2.2}^{+3.0}$ \kms deg$^{-1}$ and $x$ is
the projected abscissa on the sky.

\section{THE MASS CONTENT OF THE SCULPTOR DSPH: TWO STELLAR COMPONENTS MODELING}\label{sec:dispersion}

\begin{figure}
\begin{center}
\plotone{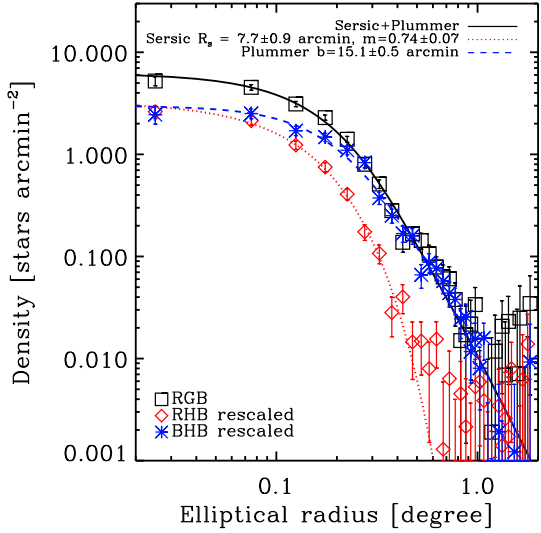}
\caption{Number surface density profile of RGB stars in Scl from
ESO/WFI photometry (squares with error-bars) overlaid to the
best-fitting two component model (solid line) given by the sum of a
Sersic (dotted line) and Plummer (dashed line) profiles. These are
obtained from the rescaled profiles that best fit, respectively, the
distribution of RHB and BHB stars (diamonds and asterisks with
error-bars, respectively) in Scl. The Galactic stellar contamination
has been subtracted from each point.}
\label{fig:density}
\end{center}
\end{figure}

Scl hosts two stellar populations with distinct spatial, metallicity
and kinematic distributions (T04 and references therein).  Our
spectroscopic sample of RGB stars shows that MR ([Fe/H]$>-1.5$) stars
are less extended spatially and have colder kinematics than MP
([Fe/H]$<-1.7$) stars\footnote{We exclude the region
$-1.7<$[Fe/H]$<-1.5$ to limit kinematic contamination between the two
populations.}.  MR and MP RGB stars appear to trace, respectively, the
Red Horizontal and Blue Horizontal Branch populations (RHB and BHB,
respectively), as shown in Figure~\ref{fig:density}. The number
surface density profile of RGB stars from our photometry is well
approximated by a two-component fit, where each component is given by
the rescaled best-fitting profile derived from our photometry for the
RHB and BHB stars. 

\subsection{Observed Velocity Dispersion Profiles}

In the following we derive the observed velocity dispersion profiles
of MR and MP RGB stars and use them jointly to constrain the Scl mass
content. However, we first need to take into account the presence of
MW interlopers which might contaminate our sample and affect the
dynamical modeling.  The probability of a MW interloper within the
3$\sigma$ kinematic sample increases with the projected distance from
the Scl center as a consequence of the decreasing ratio of the stellar
density of Scl versus the MW.  As we cannot directly weed out these
interlopers using our photometry or spectra we take them into account
in a statistical way using a maximum likelihood method.

The probability $P(v_i)$ of observing a star with a velocity $v_i$ and error $\sigma_i$ at a projected distance 
from the center $R_1\!< R <\!R_2$, where $R_1$ and $R_2$ are a generic inner and outer radius defining an 
elliptical annulus (distance bin), is:
\begin{equation}
P(v_i \mid \overline{v}, \sigma)=\frac{N_{\rm MW}}{N_{\rm T}} f_{\rm MW} + 
\frac{N}{N_{\rm T}} \frac{e^{- \frac{(v_i- \overline{v})^2}{2 ( \sigma^2+\sigma_i^2)}}}{\sqrt{2 \pi (\sigma^2 + \sigma_i^2)}}.
\end{equation}
$N_{\rm MW}$ and $N$ are the expected number of MW and Scl RGB stars in a distance bin 
($N_{\rm T}=N_{\rm MW}+N$). $f_{\rm MW}$ is the velocity distribution of MW stars, which 
we assume does not change across the face of Scl, and is  
derived from the Besan\c con model \citep{robin2003} selecting stars along the l.o.s.\ and with magnitudes and colors 
similar to the Scl RGB stars. We assume that the Scl 
velocity distribution is a Gaussian whose peak velocity $\overline{v}$ and dispersion 
$\sigma$ (the quantities we want to derive) are allowed to vary with projected radius. 
We derive the normalization factors, $N_{\rm MW}/N_{\rm T}$ and 
$N/N_{\rm T}$ directly from 
the observed RGB surface density profile and relative foreground density. 
To estimate the fraction of MW interlopers in the MR and MP sub-samples we simply count 
how many stars with velocities $ < v_{\rm sys} - 3 \sigma$ 
(i.e. the non-membership region more populated by foreground stars) are classified as MR and as MP on the basis of their CaT derived [Fe/H] value.  The likelihood of observing a set of velocities $v_i$ with $i=1,...,N$ is 
$L(v_1, ..., v_N \mid \overline{v}, \sigma)= \prod_{i=1}^N  P(v_i)$.
We maximize the likelihood function in each distance bin and find 
the corresponding best-fitting $\overline{v}(R)$ 
and $\sigma(R)$. The errors are determined from the 
intervals corresponding to 68.3\% probability.

\begin{figure}
\begin{center}
\plotone{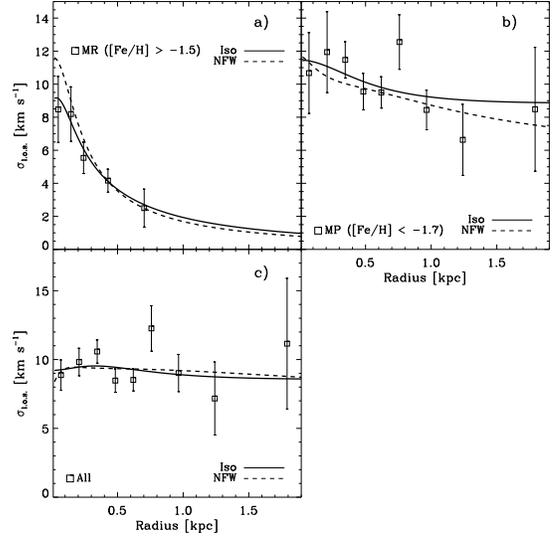}
\caption{l.o.s.\ velocity dispersion profile (squares with errorbars), from rotation-subtracted GSR velocities, 
for the MR (a), MP (b) and all (c) RGB stars in Scl. The lines show the best-fitting 
pseudo-isothermal sphere (solid) and NFW model (dashed) in the hypothesis of $\beta=\beta_{\rm OM}$. 
 Panel c) 
shows that the best-fitting pseudo-isothermal sphere with $\beta=\beta_{\rm OM}$ 
(solid) and the NFW model with $\beta=$const (dashed) are statistically indistinguishable.} 
\label{fig:sigmalos}
\end{center}
\end{figure}

The kinematics of the Scl MR and MP RGB stars 
are clearly different (Figure~\ref{fig:sigmalos}a,b): the  l.o.s.\ velocity 
dispersion profile of MR stars declines from $\sim$9 \kms in the center to 
$\sim$2 \kms at projected radius $R =$ 0.5 deg, while MP stars are kinematically hotter and 
exhibit a constant or mildly declining 
velocity dispersion profile.
  
\subsection{Predicted Velocity Dispersion Profile}

The l.o.s.\ velocity dispersion predicted by the Jeans equation for a spherical system in absence of 
net-streaming motions\footnote{We checked that the assumptions of sphericity and absence of streaming motions have a 
negligible effect on the results: the observed l.o.s.\ velocity dispersion profiles derived adopting circular 
distance bins and not subtracting rotation are consistent at the 1$\sigma$ level in each bin 
with the observed l.o.s.\ velocity dispersion profile derived adopting elliptical binning and 
by subtracting the observed rotation signal (see B07)} is \citep{binney1982}:
\begin{equation}
\label{eq:jeans_binneymamon1982}
\sigma_{\rm los}^2(R)= \frac{2}{\Sigma_*(R)} \int_R^{\infty} \frac{\rho_*(r) \sigma_{r,*}^2 \ r}{\sqrt{r^2-R^2}} 
(1 - \beta \frac{R^2}{r^2} ) dr 
\end{equation}
where $R$ is the projected radius (on the sky), $r$ is the 3D radius. The l.o.s.\ velocity dispersion depends on: 
 the mass surface density $\Sigma_*(R)$ and mass density $\rho_*(r)$
of the tracer, which in our case are the MR and the MP RGB stars; 
the tracer velocity anisotropy $\beta$, defined as $\beta = 1 -\  \sigma_{\theta}^2 / \sigma_r^2$, which we 
allow to be different for MR and MP stars; the radial velocity dispersion $\sigma_{r,*}$ for the specific component, 
which depends on the total mass distribution (for the general solution see Battaglia et al.~\citeyear{battaglia2005}).

We consider two DM mass models: a pseudo-isothermal sphere, typically
 cored, \citep[see][]{battaglia2005}, and an NFW profile, cusped \citep{nfw1996}.  Since the contribution of the stars to the total
 mass of the system is negligible for reasonable stellar M/L ratios,
 we do not consider it further. As $\beta$ is unknown we explore two
 hypotheses: a velocity anisotropy constant with radius, and an
 Osipkov-Merritt (OM) velocity anisotropy \citep{osipkov1979,
 merritt1985}.  For the latter profile, the velocity anisotropy is
 $\beta= r^2/(r^2 + r_a^2)$ where $r_a$ is the anisotropy radius.

\subsection{Results from the Two-Components Mass Modeling}
We explore a range of core radii $r_c$ for the pseudo-isothermal
sphere ($r_c\!=\!0.001, 0.05, 0.1, 0.5, 1$ kpc) and a range of
concentrations $c$ for the NFW profile ($c\!=\!20, 25, 30, 35$). By fixing
these, each mass model has two free parameters left: the anisotropy and the DM halo mass
(enclosed within the last measured point for the isothermal sphere, at
1.3 deg $=$ 1.8 kpc assuming a distance to Scl of 79 kpc, see Mateo
1998; the virial mass for the NFW model).  We
compute a $\chi^2$ for the MR and MP components separately ($\chi_{\rm
MR}^2$ and $\chi_{\rm MP}^2$, respectively) by comparing the various
models to the data. The best-fitting model is obtained by minimizing
the sum $\chi^2= \chi_{\rm MR}^2 + \chi_{\rm MP}^2$. We quote as
errors in the individual parameters the projections of the $\Delta
\chi^2=3.53 $ region (corresponding to the region of 68.3\% {\it
joint} probability for a three free parameters $\chi^2$ distribution).

The models with constant velocity anisotropy yield poor fits for both
a cored and a cusped profile.  The best-fitting models in the
hypothesis of an OM velocity anisotropy are shown in
Figure~\ref{fig:sigmalos}a,b and provide very good fits for both the
MR and MP components, simultaneously.  We find that a
pseudo-isothermal sphere with $r_c=$0.5 kpc, $M(< 1.8 \mathop{\rm
kpc}) = 3.4 \pm 0.7 \times 10^8$ \sm, $r_{\rm a, MR}=
0.2_{-0.15}^{+0.1}$ kpc and $r_{\rm a, MP}= 0.4_{-0.2}^{+0.3}$ kpc
gives an excellent description of the data ($\chi_{\rm min}^2= 6.9$,
with $\chi_{\rm MR}^2= 0.6$ and $\chi_{\rm MP}^2= 6.3$). This gives an
M/L ratio for Scl within 1.8 kpc of 158$\pm$33 (M/L)$_{\odot}$, an
order of magnitude larger than previous estimates
\citep{queloz1995}. Also an NFW model is statistically consistent with
the data ($c=20$, virial mass $M_v= 2.2_{-0.7}^{+1.0} \times 10^9$
\sm, $r_{\rm a, MR}= 0.2 \pm 0.1$ kpc and $r_{\rm a, MP}=
0.8_{-0.4}^{+2.0}$ kpc, $\chi_{\rm min}^2= 10.8$, with $\chi_{\rm
MR}^2= 4.2$ and $\chi_{\rm MP}^2= 6.6$), but tends to over-predict the
central values of the MR velocity dispersion, and this tendency is
accentuated for larger concentrations. This model gives a mass within
the last measured point of $\sim 2.4_{-0.7}^{+1.1} \times 10^8$ \sm,
which is consistent with the mass predicted by the best-fitting
isothermal sphere model.

Fig.~\ref{fig:sigmalos}c clearly shows the mass-anisotropy degeneracy
present when modeling Scl as a single stellar component system: both a
cored profile with OM velocity anisotropy and an NFW model with
constant velocity anisotropy give a very good fit to the observed
l.o.s.\ velocity dispersion profile, and they are indistinguishable.

\section{Discussion and conclusions}
We have explored the kinematic status and mass content of the Scl dSph using
 accurate line-of-sight velocities and CaT [Fe/H] measurements from VLT/FLAMES spectra
 of 470 Scl probable members. The large spatial coverage and statistics of this data-set, 
combined with the accurate velocity measurements, allowed us for the first time to detect 
in the stellar component of a dSph a statistically significant velocity gradient 
likely due to intrinsic rotation. This gradient is 
$7.6_{-2.2}^{+3.0}$ \kms deg$^{-1}$ along the Scl projected major axis. 

The presence of rotation in dSphs could support scenarios in which the progenitors of these objects 
were rotationally supported, disky systems, 
tidally stirred by the interaction with the MW \citep[e.g.][]{mayer2001}. The 
efficiency of the transformation from this kind of systems 
into dSphs, which are spheroidal in shape and have a pressure supported kinematics, 
is dependent on the eccentricity of the orbit and 
pericentric distance with respect to the host galaxy. It is likely that objects which have not been efficiently 
stirred could retain 
some of their initial kinematic characteristics, such as for example rotation. 

Under the hypothesis that Scl is not tidally disrupted we have
determined its mass using the Jeans equations, which assume dynamical
equilibrium.  The combined fit of MR and MP stars allows us to to
partly relieve the mass-anisotropy degeneracy present in modeling Scl
as a single stellar population and to exclude the models with constant
velocity anisotropy (Battaglia et al., in preparation).  The best-fit
is given by a cored profile of $r_c=$0.5 kpc and mass $M(< 1.8 \mathop{\rm kpc}) = 3.4
\pm 0.7 \times 10^8$ \sm, assuming that the velocity
anisotropy follows an OM model.  This result is in a 1.5$\sigma$
agreement with the results of \citet{strigari2007}, who performed a
careful comparison between different systems and datasets, measuring
the mass of each galaxy within the same distance from the center. Our
result makes Scl more massive than previously thought, and this does
not support the conjecture that all dSphs have a common mass scale
\citep{gilmore2007}.

\acknowledgments 
We thank L.Sales for kindly providing the orbits shown in Fig.1.

{\it Facilities:} \facility{VLT:Kueyen (FLAMES)}.

\end{document}